\documentclass[prb,superscriptaddress,aps,letterpaper,twocolumn,nofootinbib]{revtex4}

\usepackage{graphicx}
\usepackage{amsfonts}
\usepackage{amsmath}
\usepackage{subfigure}

\begin{document}

\title{Optimized Monotonic Convex Pair Potentials Stabilize Low-Coordinated Crystals}

\author{\'E. Marcotte}
\affiliation{Department of Physics, Princeton University, Princeton, New Jersey 08544, USA}

\author{F.H. Stillinger}
\affiliation{Department of Chemistry, Princeton University, Princeton, New Jersey 08544, USA}

\author{S. Torquato}
\affiliation{Department of Physics, Princeton University, Princeton, New Jersey 08544, USA}
\affiliation{Department of Chemistry, Princeton University, Princeton, New Jersey 08544, USA}
\affiliation{Princeton Center for Theoretical Science, Princeton University, Princeton, New Jersey 08544, USA}

\begin{abstract}
We have previously used inverse statistical-mechanical methods 
to optimize isotropic pair interactions with multiple extrema to yield
low-coordinated crystal classical ground states 
(e.g., honeycomb and diamond structures)
in $d$-dimensional Euclidean space $\mathbb{R}^d$.
Here we demonstrate the counterintuitive result that no extrema are required
to produce such low-coordinated classical ground states. 
Specifically, we show that monotonic convex pair potentials can be optimized
to yield classical ground states that are the square and 
 honeycomb crystals in $\mathbb{R}^2$ over a non-zero number density 
range. Such interactions may be feasible to achieve experimentally
using colloids and polymers.
\end{abstract}

\maketitle

The forward approach of statistical mechanics focuses on finding the structure and macroscopic
properties of many-particle systems with specified interactions. This approach has led to the discovery of rich and complex many-particle
configurations.\cite{Watzlawek1999,Gottwald2004,Hynninen2006}
The power of the inverse 
statistical-mechanical approach is that it can be
employed to design interactions that yield a targeted many-particle
configuration with desirable bulk physical properties.\cite{To09}
This work continues
our general program to use inverse approaches to optimize pair interactions
to achieve novel targeted classical ground-state configurations in
$d$-dimensional Euclidean space $\mathbb{R}^d$.
In particular, we have found optimized  pair interactions that
yield low-coordinated crystal classical ground states (e.g., square
and honeycomb crystals \cite{Rechtsman2006a} in $\mathbb{R}^2$, and
simple cubic \cite{Rechtsman2006b} and diamond \cite{Rechtsman2007a} crystals
in $\mathbb{R}^3$), materials with negative thermal expansion,
\cite{Rechtsman2007b} negative Poisson's ratio \cite{Rechtsman2008} and
designed optical properties.\cite{Batten2008}
We envision using colloids and/or polymers to realize such
designed potentials because one can tune their 
interactions.\cite{Lindenblatt2001,Manoharan2003,Valignat2005,To09}

Earlier uses of the inverse approach \cite{Rechtsman2006a} did
not regard experimental feasibility as a constraint. These investigations
allowed a largely unconstrained class of spherically symmetric pair potentials.
In some instances in which the goal
was to target low-coordinated crystal ground states, 
it was shown that only a few potential wells were 
required, \cite{Rechtsman2006a, Rechtsman2007a} which nonetheless may 
be difficult to realize experimentally. If purely repulsive monotonic
pair potentials existed that could achieve
unusual ground states, they would be easier to produce
experimentally. However, encoding information in 
monotonic potentials to yield low-coordinated ground-state configurations
in Euclidean spaces is highly nontrivial. Such potentials 
must not only avoid close-packed (highly coordinated) competitors
but crystal configurations that are infinitesimally close
in structure (very slight deformations of the targeted low-coordinated
crystal), which is a great challenge to achieve \emph{theoretically}
while maintaining the monotonicity property.

In this Letter, we use a {\it modified} inverse approach  to obtain
monotonic {\it convex} potentials whose ground states in $\mathbb{R}^2$ are
either the square lattice or  honeycomb crystal.\cite{Cohn2009} 
Thus, our work is a theoretical proof of concept that monotonic convex potentials 
can stabilize low-coordinated crystals.
\footnote{
We consider monotonic potentials that are also convex
because this class of  interactions may be amenable
to rigorous analysis. Indeed, in Ref.~14, 
Cohn and Kumar have rigorously constructed potentials that stabilize
unusual targeted configurations on the surface of a $d$-dimensional sphere
using only monotonic convex pair potentials. Restriction to compact spaces made 
their problem much easier to solve because their pair potentials had 
compact support set by the sphere radius. Nonetheless,
their results are suggestive that similar proofs
can be constructed in $\mathbb{R}^d$.
}
We begin by describing briefly the procedure that we employ  to optimize 
the monotonic convex potentials for the targeted low-coordinated crystals
aided by {\it generalized coordination} functions; see Ref.~15. 
This is followed by an analysis of their stability characteristics.

We consider the total potential energy $\Phi_N(\mathbf{r}^N)$ of a configuration
$C$ of $N$ particles with positions $\mathbf{r}_1$, $\mathbf{r}_2$, ...
$\mathbf{r}_N$ to be given by a sum of pairwise terms:
\begin{equation}
\Phi_N\left( \mathbf{r}^N \right) = \sum_{i < j} v(r_{ij}),
\end{equation}
where $v(r)$ is the isotropic pair potential and
$r_{ij} = | \mathbf{r}_i - \mathbf{r}_j |$.
For a targeted configuration $C^*$ to be a ground state associated with a potential $v$, 
the total potential energy needs to satisfy the following property:
\begin{equation}
\Phi_N(v,C^*) \le \Phi_N(v,C) \ \ \forall C. \label{eqn:ineq_all_struct}
\end{equation}
By expressing the potential $v$ as a function of $M$ parameters
$a_1$, ..., $a_M$, i.e.,
$v \equiv v(a_1, \dots, a_M)$, it should be possible to find the optimized potential by varying the parameters
until inequality (\ref{eqn:ineq_all_struct}) is satisfied for all possible
configurations $C$.
Due the uncountably infinite number of possible configurations, it is
impossible to check them all. Instead, we restrict ourselves to a subset of
them, which we call the {\it competitor configurations} $\mathbf{C}$.
This allows us to redefine the problem as an optimization,\cite{Rechtsman2006a}
where the objective is to maximize the energy
difference between the targeted configuration $C^*$ and its closest
competitors. This is done by introducing a utility variable $\Delta$
which is to be maximized while satisfying the following constraints:
\begin{equation}
\Phi_N\left(v(a_1, \dots, a_M),C^*\right) \le \Phi_N\left(v(a_1, \dots, a_M),C\right) - \Delta \ \ \forall C \in \mathbf{C}.
\end{equation}
For a fixed potential $v$, the utility variable $\Delta$ can only be as large
as the smallest energy difference, $\Phi_N\left(v,C\right) - 
\Phi_N\left(v,C^*\right)$, between a competitor and the targeted configuration.
Since the functional form $v$ is allowed to vary, the
optimization procedure will find the potential that maximizes the energy
difference between the targeted configuration and its closest competitor.

For a given targeted configuration $C^*$, we begin with a  
competitors set $\mathbf{C}$
that only includes the triangular lattice. A trial pair potential
$v_1$ is optimized using that set, and its putative ground state $C_1$ is 
computed using the Metropolis Monte Carlo algorithm. Namely, 
we attempt to determine the ground state for the trial potential $v_1$
by generating an initial configuration from a Poisson point process. 
This configuration
is then slowly annealed using the Metropolis  scheme down
to zero temperature. Since this algorithm cannot guarantee that the
obtained configuration is the ground state, we repeat the procedure multiple
times and keep the lowest-energy configuration
as the {\it trial}  ground state $C_1$ of the trial potential $v_1$.

If the energy of $C_1$ is
lower than that of $C^*$ for the trial potential, it proves that $C^*$ is
not the ground state of $v_1$. To discriminate against potentials with
$C_1$ as their ground states, we add $C_1$ to the list of competitors
$\mathbf{C}$, before optimizing a new trial potential $v_2$. This procedure
is repeated until we are confident that the trial-potential ground state is
indeed $C^*$, at which point we have found our optimized potential.
This method is adapted from the one presented by 
Cohn and Kumar,\cite{Cohn2009} with the difference being that we
only add to $\mathbf{C}$ configurations that have lower energies than
$C^*$ for a given trial potential.

We restrict ourselves to $v(r)$ that are sums of $M=12$ negative powers of $r$,
with a cutoff at $r = R > 0$:
\begin{equation}
v(r) \equiv \left\{
	\begin{array}{ll}
	\displaystyle \sum_{i=1}^M \frac{a_i}{r^i} & r \leq R , \\
	0 & r > R .
	\end{array}
	\right. \label{eqn:pot_form}
\end{equation}
Additionally, we only consider continuous potentials whose first and second
derivatives are also continuous at the cutoff. These two conditions 
guarantee that the interaction forces
$-\nabla_{\mathbf{r}^N} v$ are continuous and that the phonon spectra can be
calculated, respectively. Furthermore, a scale is imposed by setting
$v(r=1) = 1$.

\begin{figure}[htp]
	\centering
	\subfigure[]{
		\includegraphics[scale=0.25]{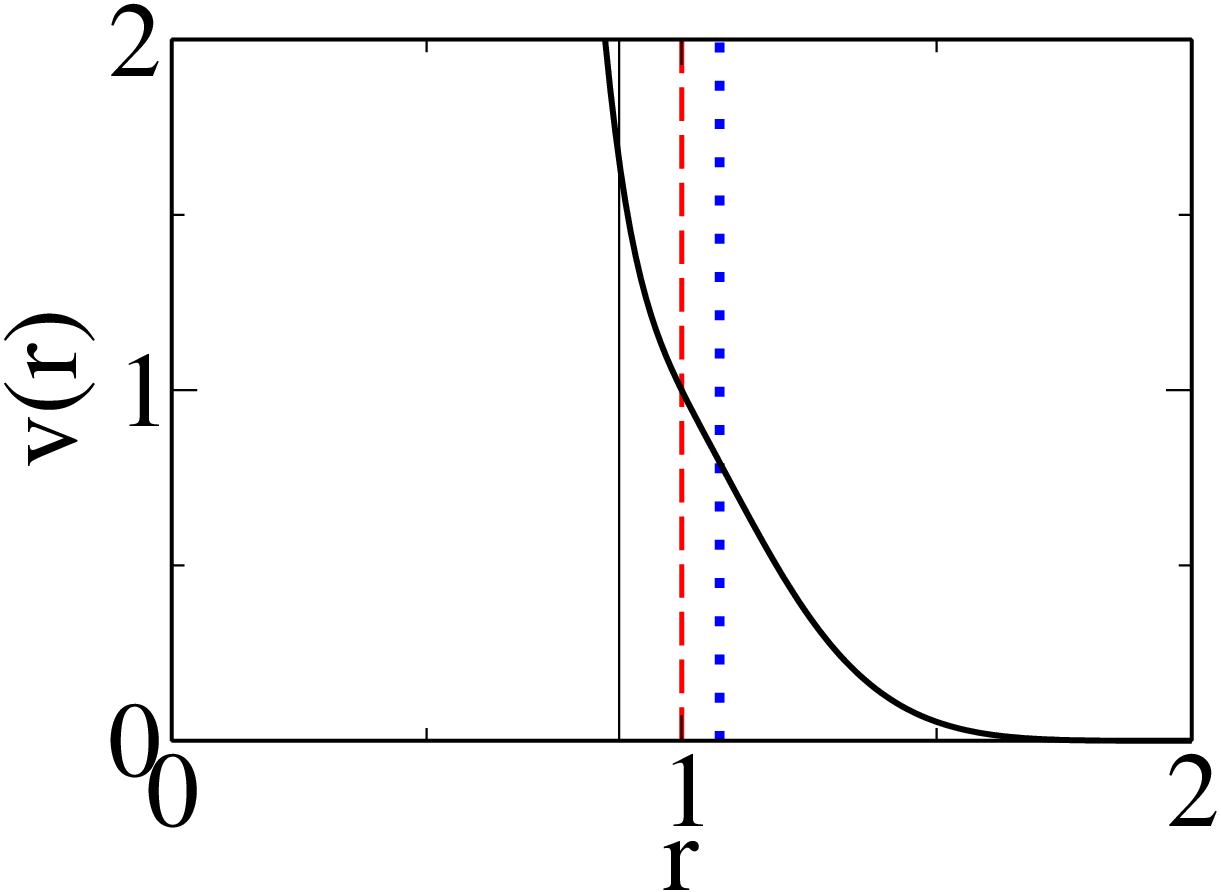}
		\label{fig:square_pot}
	}

	\subfigure[]{
		\includegraphics[scale=0.2]{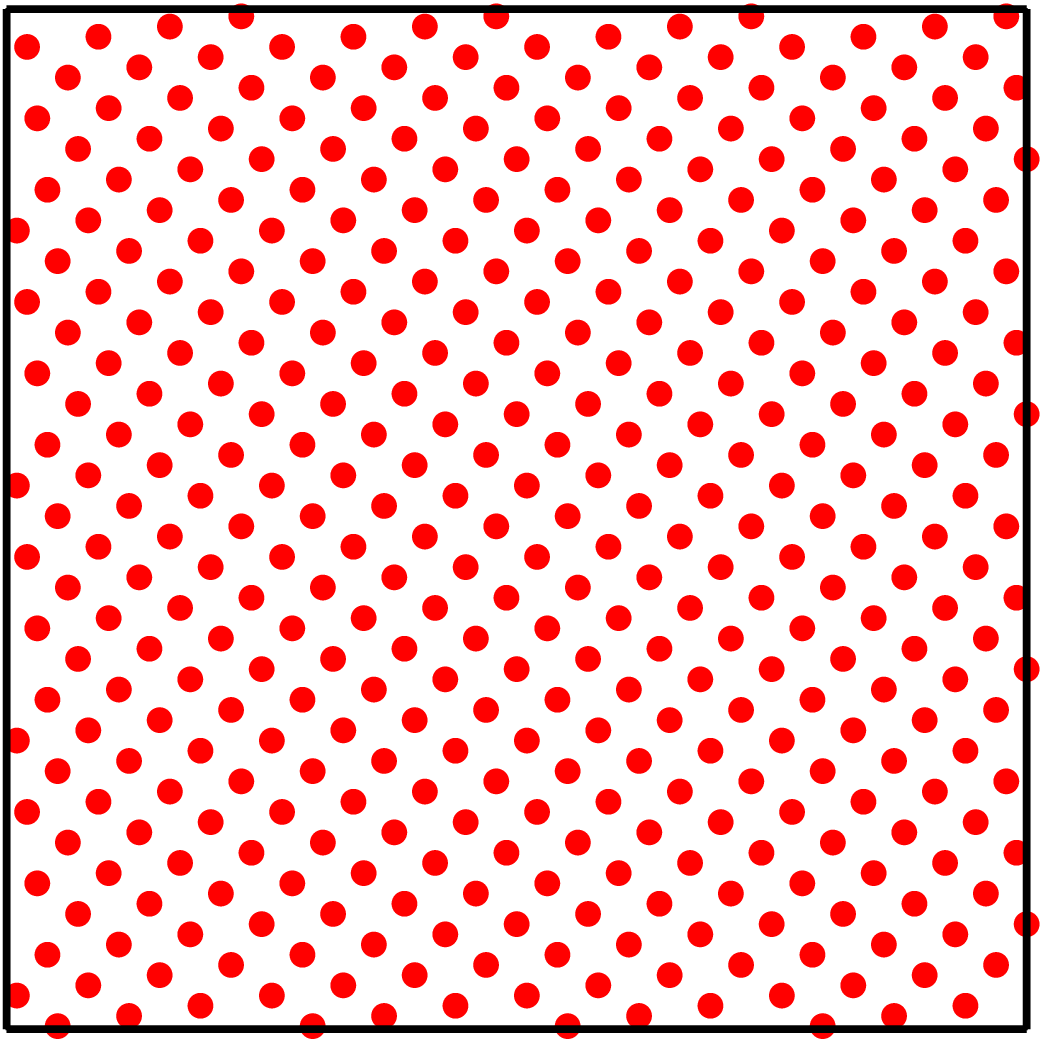}
		\label{fig:square_gs}
	}
	\caption{(Color online) \subref{fig:square_pot} Optimized convex pair
	potential targeting the square lattice. The potential  $v(r)$ at unit
	distance (where $r$ is measured in terms of the nearest-neighbor distance) 
    is taken to be unity. The vertical lines represent the 
	nearest-neighbor distances for the honeycomb crystal (black solid),
	the square lattice (red dashed) and  triangular lattice (blue dotted)
	at a  number density of unity ($\rho = 1$).
	\subref{fig:square_gs} The optimized potential ground state, obtained by
	slowly annealing the systems, starting from a fluid. The annealing was 
	performed in a $20 \times 20$ box containing $400$ particles
    under periodic boundary conditions. 
	For illustration purposes, the point particles are shown to have 
	finite sizes.}
	\label{fig:square_both}
\end{figure}

The first low-coordinated crystal configuration  to be targeted with our simulated-annealing 
optimization method is the square lattice with a nearest-neighbor
distance of unity subject to the condition that $a_i \in [-1000,1000]$.  The number density for such a configuration is unity
($\rho = 1$). The optimization procedure is restricted to monotonic convex
pair potentials that are zero beyond a cutoff distance $r = R = 2$.
We find the following optimized pair potential:
\begin{equation}
v(r) = \left\{
	\begin{array}{ll}
	\displaystyle \left( \frac{28.424}{r} - \frac{245.756}{r^2} + \frac{786.742}{r^3} - \frac{1000}{r^4} \right. \\
	\displaystyle - \frac{24.043}{r^5} + \frac{1000}{r^6} -\frac{47.967}{r^7} -\frac{1000}{r^8} \\
	\displaystyle \left. + \frac{64.527}{r^9} + \frac{1000}{r^{10}} -\frac{712.166}{r^{11}} + \frac{151.240}{r^{12}} \right) & r \leq 2 , \\
	0 & r > 2 ,
	\end{array}
	\right.
\label{eqn:square_pot}
\end{equation}
which is plotted in Fig. \ref{fig:square_pot}. To confirm that the ground state
of potential (\ref{eqn:square_pot}) is indeed the square lattice, we performed
multiple simulated-annealing calculations. All of them resulted in either
square lattices (as shown in Fig. \ref{fig:square_gs}) or
slightly deformed square lattices, whose energies were always higher than
that of the perfect square lattice.
We use $M=12$ terms in our potential because
higher $M$ cause numerical instabilities and lower $M$ result in
potentials that only weakly discriminate against competitors.
 Importantly, the potential function (\ref{eqn:square_pot}) is only one example 
within a large class of functions that could be optimized to 
stabilize the square lattice.

\begin{figure}[htp]
	\centering
	\subfigure[]{
		\includegraphics[scale=0.25]{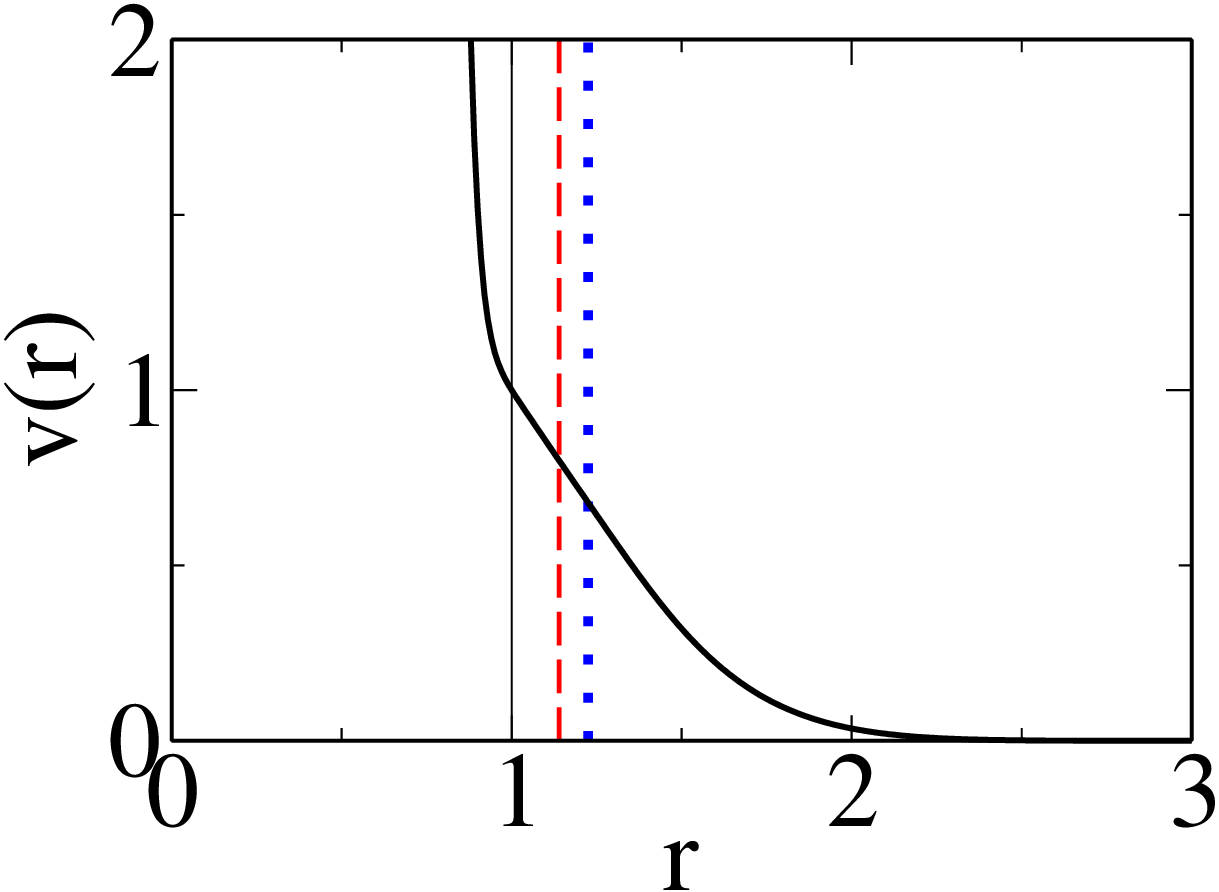}
		\label{fig:honey_pot}
	}

	\subfigure[]{
		\includegraphics[scale=0.2]{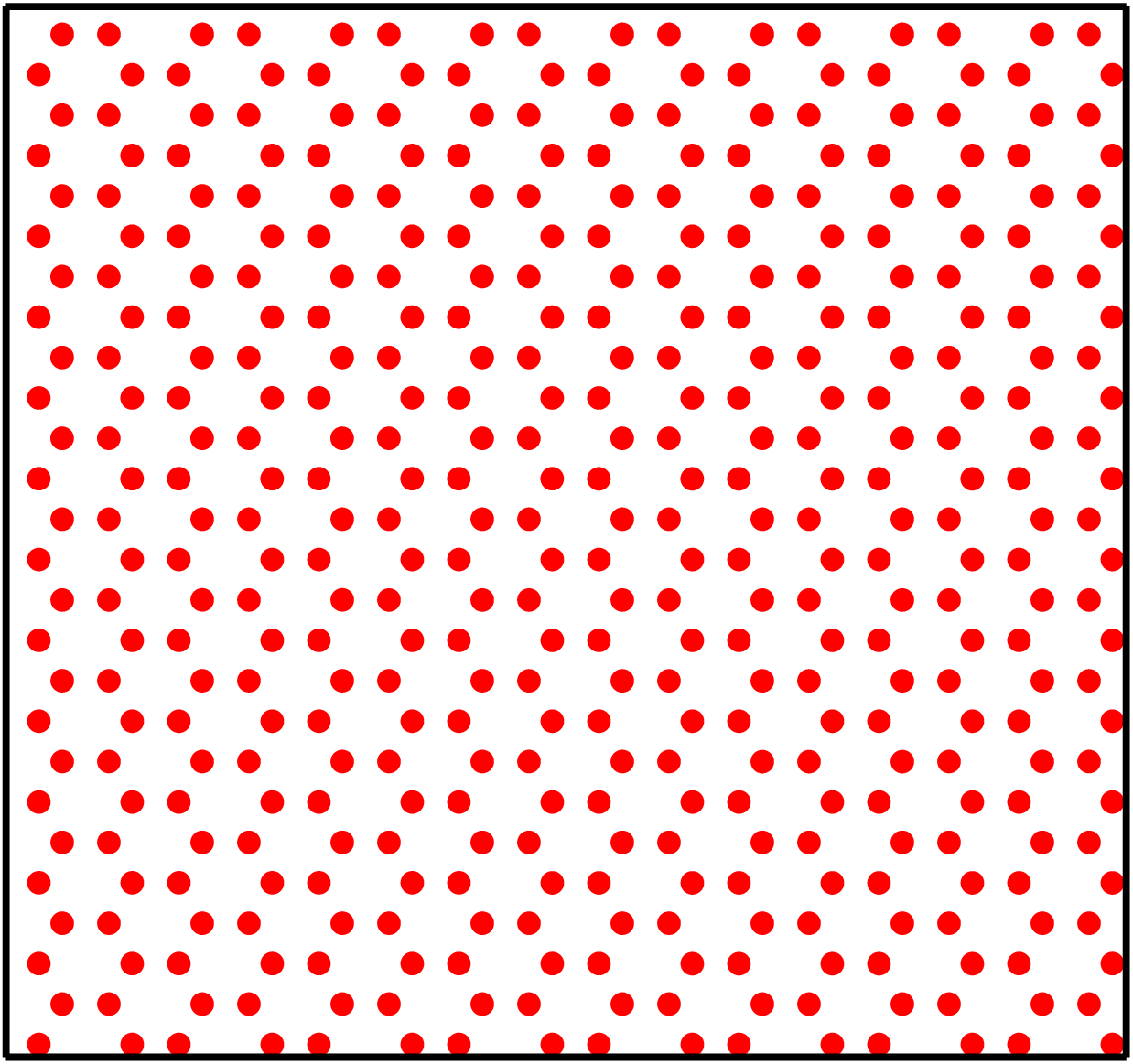}
		\label{fig:honey_gs}
	}
	\caption{(Color online) As in Fig. \ref{fig:square_both}, except that it
	is for the optimized potential targeting the honeycomb crystal.
	In \subref{fig:honey_pot}, the nearest-neighbor distances are calculated
	for $\rho = 4 / 3 \sqrt 3$;
	and in \subref{fig:honey_gs} we use $416$ particles
        in a periodic box with dimensions $24 \times 13 \sqrt 3$. }
	\label{fig:honey_both}
\end{figure}

To see intuitively why the purely repulsive potential (\ref{eqn:square_pot})
succeeds in stabilizing the square lattice consider the interactions due to
the first and/or second coordination shells,
the main contributors to the total energy. For the
square, triangular and honeycomb crystals (with $\rho = 1$),
the first coordination shell contributions to
twice the total energies per particle $u$
[$u = \sum_{j \neq 1} v(r_{1j})$ for the studied crystals]
are respectively given by
$4 \times v(1) = 4$,
$6 \times v(1.075) = 4.762$ and
$3 \times v(0.877) = 4.963$ (these can be compared to the entire function $u$,
which is respectively
$4.456$, $4.764$ and $5.236$). We see that the lower value of the
optimized potential (\ref{eqn:square_pot}) for the triangular lattice at its
nearest-neighbor distance  is not enough to compensate for the higher
coordination number compared to the square lattice (six instead of only four).
Thus, the relatively slow decrease of the potential (\ref{eqn:square_pot})
around  $r = 1$ allows it to favor configurations with low coordination,
even if it means having closer nearest neighbors.
The lower-coordinated honeycomb crystal is also discriminated against,
due the large increase of $v(r)$ for $r < 1$.

However, there are more subtle configurations that have to be
discriminated against beside the aforementioned ones.
These include infinitesimally close configurations,
such as the rhombical and rectangular lattices.
The difference in $u$ between a rectangular lattice of aspect
ratio $1 + \varepsilon$ and the square lattice is equal to
$[v'(1)/2 + \sqrt{2} v'(\sqrt{2}) + v''(1)/2] \varepsilon^2
+ O(\varepsilon^3) = 2.551 \varepsilon^2 + O(\varepsilon^3)$.
The difference in $u$ between a rhombical lattice of angle
$\pi/2 - \varepsilon$ and the square lattice is
$[v'(1) + v'(\sqrt{2})/\sqrt{2} + v''(\sqrt{2})] \varepsilon^2
+ O(\varepsilon^3) = 2.110 \varepsilon^2 + O(\varepsilon^3)$.
Stabilizing the square lattice against these two very
close neighboring configurations is thus an equilibrium between
having large second derivatives at the two first coordination shells,
while having preventing the first derivative to be too negative at
these two shells.

The second targeted ground-state configuration
is the honeycomb crystal with a nearest-neighbor distance of unity
and number density  $\rho = 4 / 3 \sqrt 3$. This is a more challenging
ground state to achieve with a monotonic convex potential because
it is only trivalently coordinated.
The
optimization procedure is still restricted to monotonic convex pair potentials,
but the cutoff is set to $R = 3$. The optimized pair potential is given by
\begin{equation}
v(r) = \left\{
	\begin{array}{ll}
	\displaystyle \left( \frac{3.767}{r} - \frac{48.246}{r^2} + \frac{230.514}{r^3} - \frac{451.639}{r^4} \right. \\
	\displaystyle + \frac{56.427}{r^5} + \frac{1000}{r^6} -\frac{868.468}{r^7} -\frac{776.495}{r^8} \\
	\displaystyle \left. + \frac{1000}{r^9} + \frac{521.638}{r^{10}} -\frac{1000}{r^{11}} + \frac{333.502}{r^{12}} \right) & r \leq 3 , \\
	0 & r > 3 ,
	\end{array}
	\right.
\label{eqn:honey_pot}
\end{equation}
which is plotted in Fig. \ref{fig:honey_pot}. As for potential
(\ref{eqn:square_pot}), we confirmed that the ground state of potential
(\ref{eqn:honey_pot}) is indeed the honeycomb crystal by performing
multiple simulated-annealing calculations. Figure \ref{fig:honey_gs} shows
the result of one of those runs that converged to the honeycomb crystal.
All final configurations other than the honeycomb crystal had
higher energies than that of the perfect honeycomb, which is strong numerical
evidence that the honeycomb is indeed the ground state of
potential (\ref{eqn:honey_pot}).

As in the square-lattice case, the ability of the purely
repulsive potential (\ref{eqn:honey_pot}) to stabilize the low-coordinated
honeycomb crystal lies in its slow decrease near $r = 1$. Consequently,
the contributions to $u$
from the first coordination shells of the honeycomb, triangular and
square crystals (at $\rho = 4 / 3 \sqrt 3$), which are respectively
$3$, $4.077$ and $3.196$. The low coordination of the honeycomb crystal
thus compensate for closer neighbors. The second shell
energy contributions of the square and triangular lattices 
turn out to be larger than that for the honeycomb crystal, even if all
them are relatively small due to the rapid decrease of $v(r)$.
We have also verified that potential (\ref{eqn:honey_pot}) discriminates
against slightly sheared deformations of the honeycomb crystal, which is
consistent with our phonon analysis below.

\begin{figure}[htp]
	\centering
	\subfigure[]{
		\includegraphics[scale=0.2]{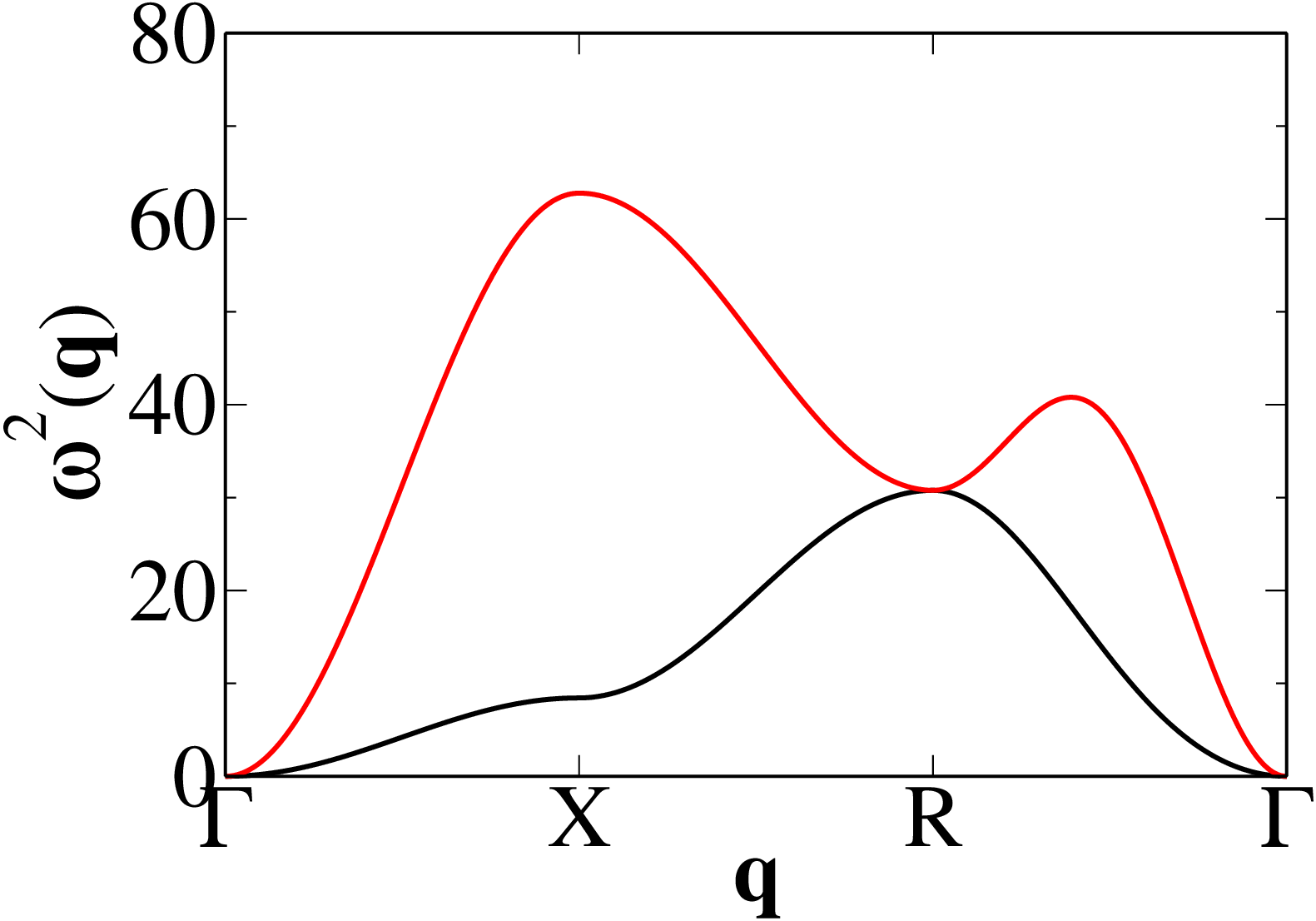}
		\label{fig:square_phonon}
	}

	\subfigure[]{
		\includegraphics[scale=0.2]{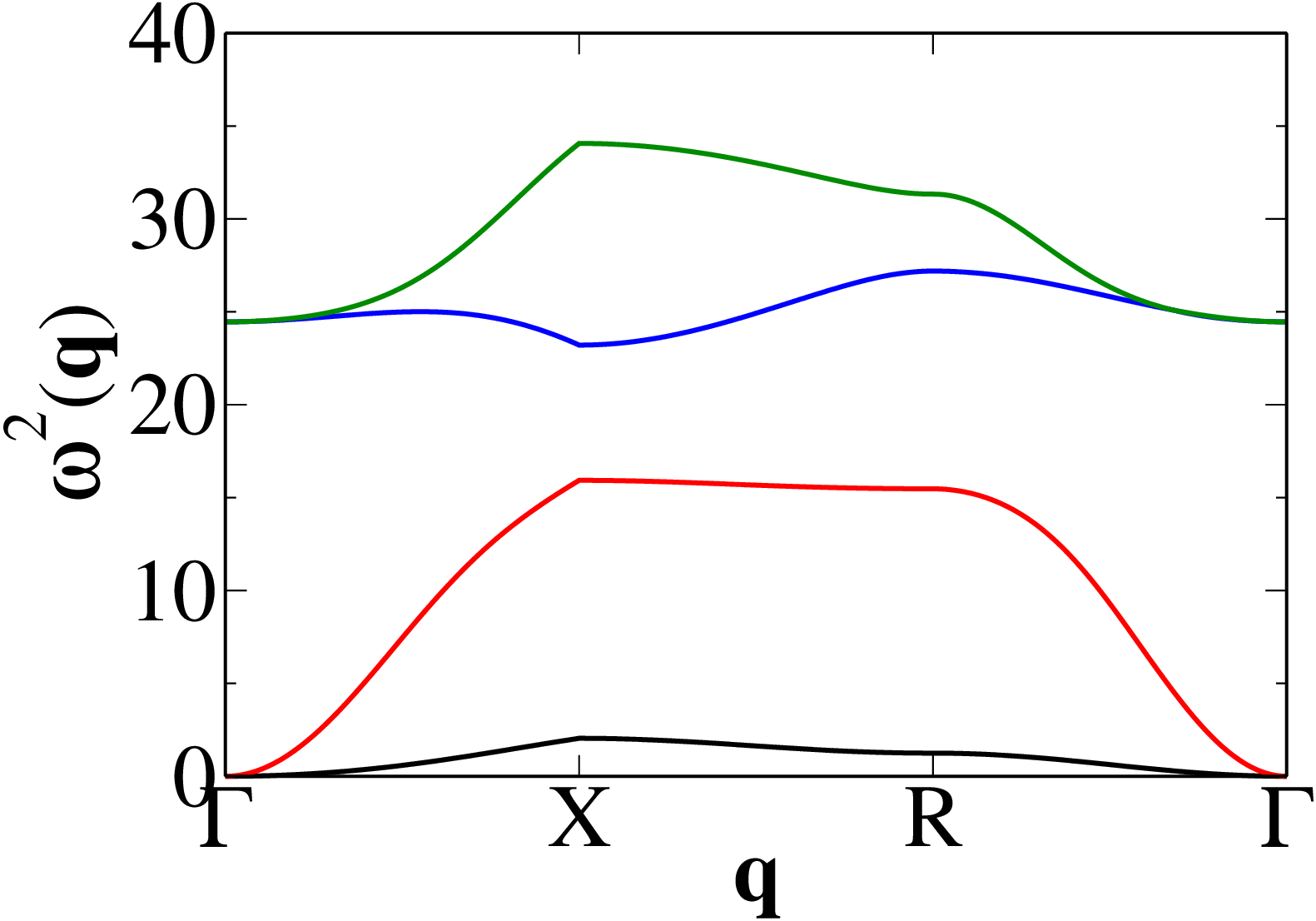}
		\label{fig:honey_phonon}
	}
	\caption{(Color online) Phonon spectra for the optimized potentials. 
	The squared phonon frequency $\omega^2$ is plotted in term of a 
	representative subset of phonon wave vectors $\mathbf{q}$.
	Note that the spectra were calculated over the entire Brillouin zone,
	and no modes with imaginary frequencies were found.
	\subref{fig:square_phonon} Phonon spectrum for the ``square-lattice''
	potential shown in Fig. \ref{fig:square_pot}.
	\subref{fig:honey_phonon} Phonon spectrum for the ``honeycomb-crystal''
	potential shown in Fig. \ref{fig:honey_pot}.}
	\label{fig:both_phonon}
\end{figure}

We also studied the phonon characteristics of potentials
(\ref{eqn:square_pot}) and (\ref{eqn:honey_pot}), i.e., the
mechanical response of the crystals to small
deformations. Figures \ref{fig:square_phonon} and \ref{fig:honey_phonon}
show the squared frequency of phonon modes as a function of their
wave vectors for the square and honeycomb crystals. The absence of any
negative squared frequency indicates that all of the modes have real frequency;
thus, the crystals are stable under small deformations.

We have further studied the stability of our optimized potentials 
by  exploring the effects of adding point defects to the crystals.
Since defects cost energy, our targeted 
ground states are stable under such local modifications. 
We also used the newly introduced
\textit{generalized coordination functions} \cite{Ma10} to
show that potentials (\ref{eqn:square_pot}) and (\ref{eqn:honey_pot}) are
part of a  large class of monotonic convex pair
potentials that stabilize the square and honeycomb crystals and thus
our potentials are robust against shape  change.
These details are given in our companion paper.\cite{Ma10}

Are the low-coordinated crystal ground states stable over a density
range around the density values for which they are designed
($\rho = 1$ for the square lattice and $\rho = 4 / 3 \sqrt 3$
for the honeycomb crystal)? We have performed both simulated-annealing ground-state
calculations and computed the phonon spectra at various densities.
We find that for potential (\ref{eqn:square_pot}), the square lattice is the ground
state for the density range $\rho \in [0.96, 1.10]$. For potential (\ref{eqn:honey_pot}),
the honeycomb crystal is the ground state for $\rho \in [0.74, 0.80]$.
Therefore, both targeted configurations can be stabilized over a non-zero
number density range, which is a desirable feature for experimental
realizations of our optimized potentials.
This property is not at all obvious for system under positive
pressure, as their are no a priori reasons why deformed lattices that maintain
a constant nearest neighbor distance, such as
the rhomboidal and rectangular lattices, are not the ground states for
densities other than the one at which the optimization was conducted.

To summarize, whether potentials exist that stabilize low-coordinated
crystal ground states in Euclidean space without
any potential wells is not at all obvious. We have shown that
potentials without wells, namely, monotonic convex repulsive pair interactions,
can produce low-coordinated ground states in $\mathbb{R}^2$, such as the
square lattice and honeycomb crystals.
Therefore, the naive expectation that purely repulsive interactions
will only lead to densely packed configurations in Euclidean space
has been disproved.

Lindenblatt \textit{et al.} \cite{Lindenblatt2001} have
fabricated so-called ``hairy
colloids''. These colloids are formed by grafting polymer chains onto the
surface of nanoscopic microgel spheres, in a matrix of polymer chains.
Manipulation of such systems offer the possibility of
mimicking the interactions that
stabilize the square and honeycomb crystals
defined by Eqs. (\ref{eqn:square_pot}) and (\ref{eqn:honey_pot}), respectively,
although such experimental realizability remains
an open fascinating question. Note that experimental feasibility  
should not  require convexity. For example, we have also shown
numerically that the square lattice can be stabilized by a monotonic non
convex potential. \cite{Ma10} In future research, we intend
to determine whether convex potentials can stabilize three-dimensional
low-coordinated structures, such as simple cubic or  diamond crystals.

This work was supported by the Office of Basic Energy Sciences,
U.S. Department of Energy under Grant No. DE-FG02-04-ER46108. 
We also acknowledge support from the Natural Sciences and Engineering
Research Council of Canada.

\end{document}